\documentclass[aps]{revtex4}
\pdfoutput=1
\usepackage{eurosym}
\usepackage{amsfonts}
\usepackage{amsmath}
\usepackage{amssymb,epsf}
\usepackage{color}
\usepackage{graphicx}
\usepackage{float}
\usepackage{caption}
\usepackage{subfig}
\usepackage{epstopdf}
\usepackage{hyperref}

\begin{document}

\title{Nonlinearly charged dyonic black holes}
\author{S. Panahiyan$^{1,2,3}$\footnote{
email address: shahram.panahiyan@uni-jena.de}}
\affiliation{$^1$Helmholtz-Institut Jena, Fr\"{o}belstieg 3, D-07743 Jena, Germany  \\
	$^2$GSI Helmholtzzentrum f\"{u}r Schwerionenforschung, D-64291 Darmstadt, Germany \\
	$^3$Theoretisch-Physikalisches Institut, Friedrich-Schiller-University Jena, D-07743 Jena, Germany}

\begin{abstract}
In this paper, we investigate the thermodynamics of dyonic black holes
in the presence of Born-Infeld electromagnetic field. We show that
electric-magnetic duality reported for dyonic solutions with Maxwell
field is omitted in case of Born-Infeld generalization. We also confirm
that generalization to nonlinear field provides the possibility of
canceling the effects of cosmological constant. This is done for
nonlinearity parameter with $10^{-33}\mbox{ eV}^{2}$ order of magnitude which
is high nonlinearity regime. In addition, we show that for small
electric/magnetic charge and high nonlinearity regime, black holes would
develop critical behavior and several phases. In contrast, for highly
charged case and Maxwell limits (small nonlinearity), black holes have
one thermal stable phase. We also find that the pressure of the cold
black holes is bounded by some constraints on its volume while hot black
holes' pressure has physical behavior for any volume. In addition, we
report on possibility of existences of triple point and reentrant of
phase transition in thermodynamics of these black holes. Finally, We show
that if electric and magnetic charges are identical, the behavior of our
solutions would be Maxwell like (independent of nonlinear parameter and
field). In other words, nonlinearity of electromagnetic field becomes
evident only when these black holes are charged magnetically and
electrically different.
\end{abstract}

\maketitle

\section{Introduction}

In the past two decades, the progresses in thermodynamics of black holes
provided us with new deep insights into physics of black holes as well
as its applications in other fields. Among these insights/applications,
one can point out to development of the $AdS/CFT$ correspondence
\cite{Maldacena}, principles of holography
\cite{Witten,Aharony}, connections to hydrodynamics
\cite{Policastro}, introduction of van der Waals like behavior
\cite{Kubiznak}, reentrant of the phase transition
\cite{Gunasekaran}, existence of the triple point
\cite{Altamirano,Momennia}, analogous heat engines
\cite{Johnson,HendiHeat} and, connection between quasinormal modes and
thermodynamical phase transition \cite{Rao,Liu}.

Among different classes of black holes and their thermodynamics, dyonic
black holes are of special interest/importance. This is because this
family of the magnetically charged static black hole solutions were
proven to have wide applications in different fields of the physics. To
name a few, one can mention: the Hall conductivity and zero momentum
hydrodynamic response functions in the context of AdS/CFT
\cite{Hartnoll}, correspondence between large dyonic black holes and
stationary solutions of the equations of relativistic
magnetohydrodynamics \cite{Caldarelli}, inducing external magnetic
field effects on superconductors \cite{Albash}, investigation of
Hall conductance, DC longitudinal conductivity
\cite{Goldstein} and paramagnetism/ferromagnetism phase transitions
\cite{Dutta,HRP}. So far, dyonic black holes were obtained in effective
\cite{DS3} and hetroctic \cite{DS4} string theories,
supergravity \cite{SGD1,SGD2}, massive gravity \cite{HRP},
gravity's rainbow \cite{PHR} and dilatonic gravity
\cite{Poletti,Hajkhalili}. Their thermodynamics were investigated in
Refs. \cite{BHD2,BHD3}. In this paper, we
investigate the thermodynamics of the dyonic black holes in the presence
of Born-Infeld (BI) nonlinear electromagnetic field.

BI theory is one of the well established theories which generalizes
Maxwell field in a nonlinear framework. The primary motivation of this
theory was to remove self energy of a point like charge in Maxwell
theory \cite{Born}. Later on, it was shown that this
electromagnetic field have features such as: the absence of shock waves,
birefringence phenomena \cite{Boillat} and electric-magnetic
duality \cite{Gibbons}. So far, different classes of black holes
in the presence of BI were constructed and investigated including: BTZ
dilatonic \cite{BTZBI}, Lovelock--Born--Infeld-scalar gravity
\cite{Meng}, massive gravity \cite{HBP}, gravity's rainbow
\cite{HBPM} and iBorn-Infeld \cite{Wang}. Previously, dyonic black
holes in the presence of the BI field were obtained in Ref.~\cite{Li} and their holographic complexity were investigated
\cite{Meng2}. In this paper, we focus more on the thermodynamics of
dyonic black holes in the presence of BI nonlinear electromagnetic
field. We study the thermal phases of these black holes, their behaviors
in different regimes, limitation on having physical solutions, existence
of phenomena such as reentrant of phase transitions and triple point.
In addition, we confirm that solutions could exhibit linearly charge
dyonic black holes' behavior if certain limits, which are not expected,
are meet.

The structure of the paper is as follows: first, we present the action
governing these black holes, obtain the metric function and investigate
the geometrical properties. Then, thermodynamical quantities are
calculated. In addition, thermodynamical phases available for these
black holes, the possibility of phase transition and restrictions for
having critical behavior are studied. Next, the conditions for having
linearly charge dyonic black holes' behavior are extracted and
discussed. The paper is concluded by some closing remarks.

\section{Black hole solutions} \label{Black hole solutions}

In this paper, we intend to construct $4$-dimensional topological black
holes in the presence of BI nonlinear electromagnetic field with dyonic
charge. To do so, we consider the following action 

\begin{equation}
\mathcal{I}=-\frac{1}{16\pi }\int d^{4}x\sqrt{-g}\left[ \mathcal{R}-2\Lambda
+L(\mathcal{F})\right] ,
\label{Action}
\end{equation}%
where $g$ is the trace of metric tensor, $\mathcal{R}$ is the scalar
curvature, $\Lambda $ is the cosmological constant and $L(\mathcal{F})$
is the Lagrangian of BI theory given by

\begin{equation}
L(\mathcal{F})=4\beta ^{2}\left( 1-\sqrt{1+\frac{\mathcal{F}}{2\beta ^{2}}}%
\right) ,  \label{LBI}
\end{equation}%
in which $\beta $ is the nonlinearity parameter. The Maxwell invariant
is $\mathcal{F}=F_{\mu \nu }F^{\mu \nu }$ where $F_{\mu \nu }=
\partial _{\mu }A_{\nu }-\partial _{\nu }A_{\mu }$ is the electromagnetic
field tensor and $A_{\mu } $ is the gauge potential. 

Variation of the action \eqref{Action} with respect to the metric tensor
$g_{\mu \nu }$ and the Faraday tensor $F_{\mu \nu }$, leads to

\begin{equation}
G_{\mu \nu }+\Lambda g_{\mu \nu }-\frac{1}{2}g_{\mu \nu }L(\mathcal{F})-%
\frac{2F_{\mu \lambda }F_{\nu }^{\lambda }}{\sqrt{1+\frac{\mathcal{F}}{%
			2\beta ^{2}}}}=0,  \label{Field equation}
\end{equation}%
\begin{equation}
\partial _{\mu }\left( \frac{\sqrt{-g}F^{\mu \nu }}{\sqrt{1+\frac{\mathcal{F}%
		}{2\beta ^{2}}}}\right) =0,  \label{Maxwell equation}
\end{equation}%
where $G_{\mu \nu }$ is the Einstein tensor. 

Since we are interested in topological black holes, we consider the
metric to be

\begin{equation}
ds^{2}=-\psi (r)dt^{2}+ \frac{dr^{2}}{\psi (r)}+r^{2}\left\{ 
\begin{array}{cc}
	d\theta ^{2}+\sin ^{2}\theta d\varphi ^{2}, & k=1 \\ 
	d\theta ^{2}+d\varphi ^{2}, & k=0 \\ 
	d\theta ^{2}+\sinh ^{2}\theta d\varphi ^{2}, & k=-1%
\end{array}%
\right..  \label{metric}
\end{equation}

The constant $k$ indicates that the boundary of $t=\mathit{constant}$ and
$r=\mathit{constant}$ can be a negative (hyperbolic), zero (flat) or positive
(elliptic) constant curvature hypersurface. 

To obtain the electric-magnetic matter field, we employ the following
gauge potential
\begin{equation}
A=h(r) dt+H(\theta)d\varphi.  \label{gaugeP}
\end{equation}%

Such gauge potential should create a \textit{radial dependent electric field} and \textit{spatial dependent magnetic field}. One can show that by using Maxwell equation \eqref{Maxwell equation}, metric \eqref{metric} and gauge potential \eqref{gaugeP}, two set of solutions are found for electric-magnetic matter field

\begin{equation}
\left\{ 
\begin{array}{cc}
h(r)=\int \frac{q_{E} \beta}{\sqrt{q_{E}^2+r^4 \beta^2}} dr, & H(\theta)=q_{M}  \\ 
 \\ 
h(r)=\int \frac{q_{E}}{r^2}\sqrt{\frac{q_{M}^2+\beta ^2 r^4}{q_{E}^2+\beta ^2 r^4}} dr, & H(\theta)=q_{M}\left\{ \begin{array}{cc}
\sin\theta, & k=1 \\ 
\theta, & k=0 \\ 
\sinh\theta, & k=-1%
\end{array}\right.
\end{array}%
\right. ,  \label{gauge}
\end{equation}
where $q_{E}$ and $q_{M}$ are integration constants related to total
electric and magnetic charges.
The first set of solutions indicates that
the magnetic part of gauge potential is constant. This results into
vanishing magnetic component of the electromagnetic tensor, hence
absence of magnetic effects in metric function. Therefore, this set of
solutions is not of interest. The second set of solutions shows that
magnetic part is not affected by nonlinearity generalization. In
contrast, the electric part is affected by nonlinearity parameter and
also magnetic charge. In general, we have two conditions for
constructing dyonic black holes in the presence of nonlinear
electromagnetic field: I) The magnetic part of the electromagnetic
tensor should be constant while the electric part is a varying one. II)
In the weak limit ($r \longrightarrow \infty \text{ or } \beta \longrightarrow
\infty $), the electric and magnetic fields should decay to Maxwellian
one. The first condition is satisfied for obtained solutions and the
second condition could be confirmed as follows

\begin{equation}
\lim_{\beta \longrightarrow \infty} h'(r)=\frac{q_{E}}{r^2}-\frac{q_{E}^3}{2 r^6 \beta^2}+O(\frac{1}{\beta^4}),
\end{equation}

Next step is calculation of metric function and testing whether it could
admit black hole solution. We use Eq. \eqref{Field equation} with given
metric \eqref{metric} and obtained gauge potential \eqref{gauge} to find
following field equations

\begin{eqnarray}
e_{tt} & = & e_{rr} = r^3 \psi'(r)+r^2 \psi(r)+\frac{2 q_{M}^2+2 \beta ^2 r^4}{\sqrt{\frac{q_{M}^2+\beta ^2 r^4}{q_{E}^2+\beta ^2 r^4}}}-2 \beta ^2 r^4+\Lambda  r^4=0,
\\[0.1cm]  
e_{\theta \theta} & = & e_{\phi \phi} = -r^3 \left(q_{E}^2+\beta ^2 r^4\right) \sqrt{\frac{q_{M}^2+\beta ^2 r^4}{q_{E}^2+\beta ^2 r^4}} \left(r \left(4 \beta ^2-\psi''(r)-2 \Lambda \right)-2 \psi'(r)\right)-4 q_{E}^2 q_{M}^2+4 \beta ^4 r^8=0,
\end{eqnarray}
where $\psi '(r)=\frac{d\psi (r)}{dr}$ and $\psi ''(r)=\frac{d^{2} \psi (r)}{dr^{2}}$. The metric function is obtained as 

\begin{eqnarray}
\psi (r) & = & 	k-\frac{m}{r}+\frac{2 \left(q_{E}^2+\beta ^2
	r^4\right)}{r^2}  \sqrt{\frac{q_{M}^2+\beta ^2 r^4}{q_{E}^2+\beta ^2 r^4}}+\frac{(2 \beta ^2 - \Lambda) r^2}{3}   -                                      \notag
\\[0.1cm] 
&   &  
\frac{4 \beta ^2 r^2 \left(6 \beta ^2 r^4 F_1\left(\frac{7}{4};\frac{1}{2},\frac{1}{2};\frac{11}{4};-\frac{r^4 \beta ^2}{q_{E}^2},-\frac{r^4 \beta ^2}{q_{M}^2}\right)+7 \left(q_{E}^2+q_{M}^2\right)
	F_1\left(\frac{3}{4};\frac{1}{2},\frac{1}{2};\frac{7}{4};-\frac{r^4 \beta ^2}{q_{E}^2},-\frac{r^4 \beta ^2}{q_{M}^2}\right)\right)}{21 q_{E} q_{M}}, \label{fr}
\end{eqnarray}
in which $m$ is integration constant related to total mass and known as
geometrical mass, and $F_{1}$ is Appell Hypergeometric function
\cite{Schlosser}.

The term $2 \beta ^{2} - \Lambda $ enables us to remove the
cosmological constant's effects. This indicates that effective behavior
of the cosmological constant (a gravitational correction) could be
canceled out by the presence of nonlinear electromagnetic field (a
matter field correction). This results into absence of observational
evidences of cosmological constant's contributions in the behavior of
such black holes. The cosmological constant is associated to vacuum
energy and according to the latest observations, its value is about
$10^{-66}\mbox{ eV}^{2}$ in natural unit. To cancel out the effects of the
cosmological constant, the nonlinearity parameter should be in order of
$10^{-33}\mbox{ eV}^{2}$. This corresponds to high nonlinearity regimes, far
away from Maxwell regime. Another interesting issue is the absence of
electric-magnetic duality. In linearly charged dyonic black holes, there
is a symmetry for swapping $E$ and $M$ integration constants
\cite{Dutta,HRP}. This is called electric-magnetic duality. Here, we see
that generalization to nonlinear electromagnetic results into vanishing
of such symmetry, hence absence of electric-magnetic duality. Finally,
it should be noted that in the limit of $q_{E} \rightarrow 0$,
$q_{M} \rightarrow 0$ and $\beta \rightarrow 0$, the metric function \eqref{fr} reduces to Schwarzschild solutions in the presence of
cosmological constant

\begin{eqnarray}
\psi (r) & = & 	k-\frac{m}{r}-\frac{ \Lambda r^2}{3}.
\end{eqnarray}

\begin{figure*}[!htbp]
	\centering
	\includegraphics[width=0.4\textwidth]{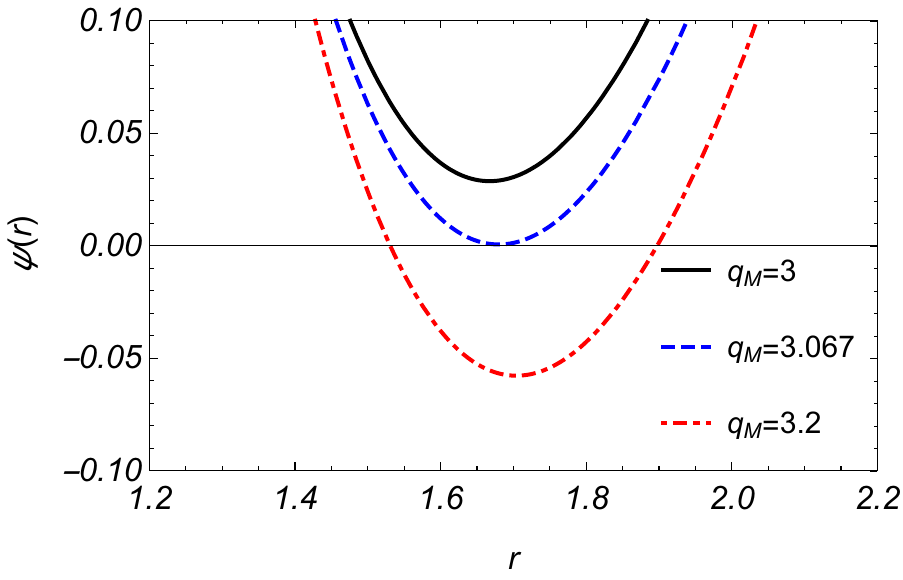}
	\includegraphics[width=0.4\textwidth]{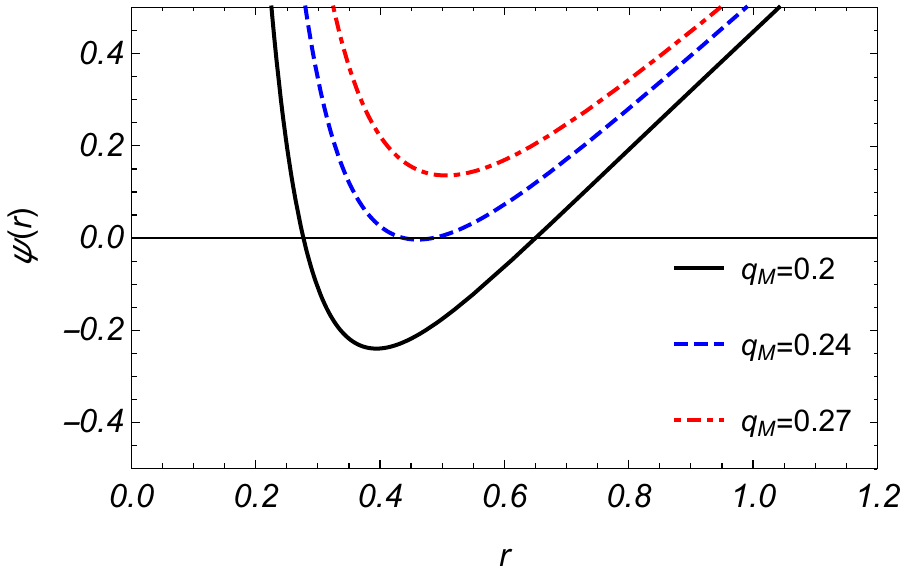}
	\caption{Metric function for $\beta =1$, $\Lambda =-1$, $q_{E}=0.5$ and $m=1$; Left panel: $k=0$, Right panel: $k=1$.} \label{Fig0}
\end{figure*}

The solutions could be interpreted as black hole ones if two conditions
are satisfied: I) Existence of event horizon for solutions. This is done by
finding positive and real valued roots for the metric function. This is
a crucial requirement that must be met under any circumstances. II)
Existence of irremovable divergency for the solutions. This is done by
investigating the curvature scalars. To address the first condition, we
use numerical method and plot diagrams in Fig.~\ref{Fig0}. Evidently,
the solutions could admit up to two roots if suitable values are
considered for different parameters. It should be noted that there is
a possibility of absence of root, hence event horizon. In this case, if
an irremovable divergency exists, it is called naked singularity. Here,
we are not interested in such solutions.

\begin{figure*}[!htbp]
	\centering
	\includegraphics[width=0.4\textwidth]{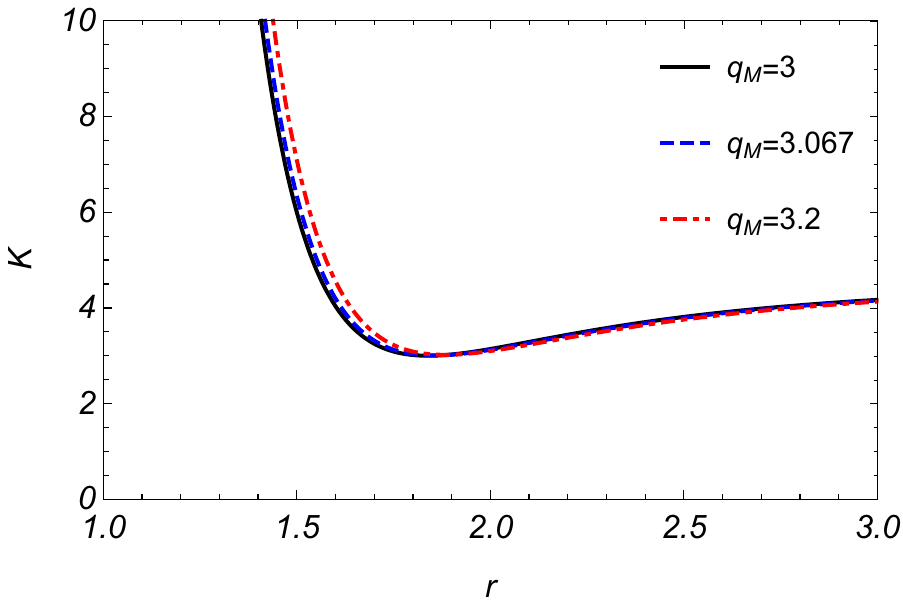}
	\includegraphics[width=0.4\textwidth]{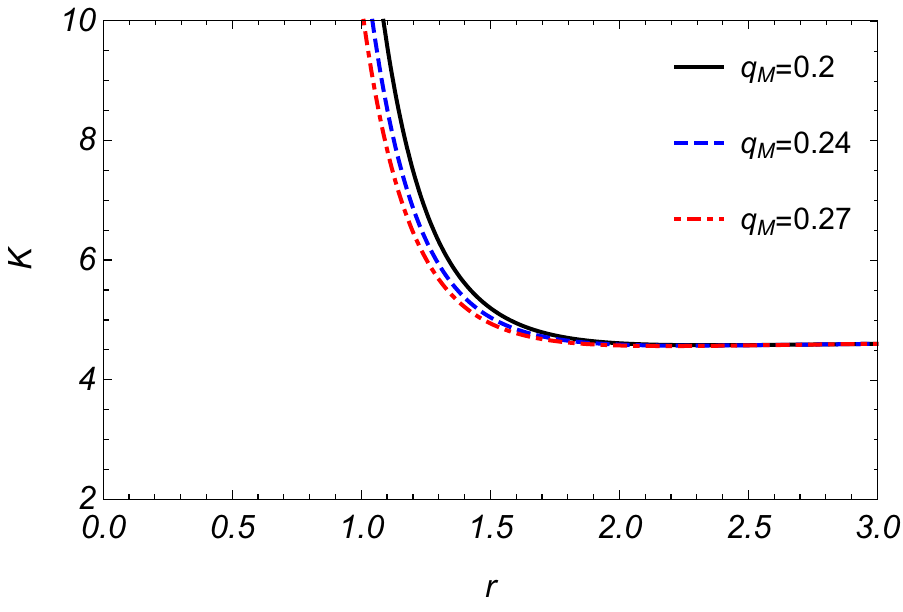}
	\caption{Kretschmann scalar for $\beta =1$, $\Lambda =-1$, $q_{E}=0.5$ and
		$m=1$; Left panel: $k=0$, Right panel: $k=1$.} \label{Fig00}
\end{figure*}

In order to investigate the second condition, we use Kretschmann scalar
given by

\begin{equation}
K=R_{\alpha \beta \gamma \delta }R^{\alpha \beta \gamma \delta}=\left( \frac{%
	d^{2}\psi\left( r\right) }{dr^{2}}\right) +\frac{4}{r^{2}}\left( \frac{d\psi\left(
	r\right) }{dr}\right) ^{2}+\frac{8} {r^{4}}\left( \psi\left( r\right)
-k\right) ^{2},  \label{Kretschmann}
\end{equation}%
where $R_{\alpha \beta \gamma \delta }$ is Riemann tensor. It is a
matter of calculation to show that limiting behavior of this scalar for
small and large radius are given by

\begin{equation}
\lim_{r\longrightarrow 0} K \longrightarrow \infty , \\
\end{equation}

\begin{equation}
\lim_{r\longrightarrow \infty } K \propto - \frac{{2\Lambda }}{3} + \frac{8 \Lambda^2}{3}.
\end{equation}
where confirm two facts: a) Our solutions contains an irremovable
divergency at the origin covered by event horizon(s). b) Solutions are
asymptotically $(A)dS$ depending on signature of the cosmological
constant. It is worthwhile to mention that Kretschmann scalar is regular everywhere outside
of the event horizon (see Fig.~\ref{Fig00}).

\section{Thermodynamics}

In this section, we calculate thermodynamical quantities and investigate
the thermodynamical behavior of the solutions.

\subsection{First law of black hole thermodynamics} 

Entropy is the first quantity that we will calculate. In general, the
calculation of the entropy depends on the gravity under consideration.
In this paper, we obtained black holes in the presence of Einstein
gravity. Therefore, we can use Hawking and Bekenstein area law
\cite{Beckenstein,Hawking} to obtain entropy as
\begin{equation}
S=\frac{1}{4}\int d^2x\sqrt{\gamma}=\frac{r_{+}^2}{4},
\label{entropy}
\end{equation}
where $\gamma $ is the induced metric on the boundary with constant
temporal and radial coordinates, {and $r_{+}$ is the event
	horizon of the black hole}. Evidently, there is no direct contributions
from matter field on entropy and the area law in its original format is
resulted (without any modification).

In next step, we use surface gravity in the following form
\cite{HawkingT} to calculate the temperature

\begin{equation}
T=\frac{\kappa }{2\pi }=\frac{1 }{2\pi }\sqrt{-\frac{1}{2}\left( \nabla _{\mu }\chi _{\nu }\right) \left(
	\nabla ^{\mu }\chi ^{\nu }\right) },
\end{equation}
where $\chi ^{\nu }$ is the time-like Killing vector. The spacetime in
this paper have a time-like Killing vector of $\chi =\partial _{t}$.
Consequently, the surface gravity is related to first order
derivation of metric function with respect to radial coordinate,
$k = \frac{1}{2}(\frac{d \psi (r)}{dr})|_{r=r_{+}}$. To obtain the final
form of the temperature, one should also calculate the geometrical mass,
$m$. We can do this by evaluating the metric function on horizon ($
\psi (r=r_{+})=0$) which results into

\begin{eqnarray}
m & = & 	kr_{+} + \frac{2 \left(q_{E}^2+\beta ^2
	r^4\right)}{r_{+}}  \sqrt{\frac{q_{M}^2+\beta ^2 r_{+}^4}{q_{E}^2+\beta ^2 r_{+}^4}}+\frac{(2 \beta ^2 - \Lambda) r_{+}^3}{3}   -                                      \notag
\\[0.1cm] 
&   &  
\frac{4 \beta ^2 r_{+}^3 \left(6 \beta ^2 r^4 F_1\left(\frac{7}{4};\frac{1}{2},\frac{1}{2};\frac{11}{4};-\frac{r^4 \beta ^2}{q_{E}^2},-\frac{r^4 \beta ^2}{q_{M}^2}\right)+7 \left(q_{E}^2+q_{M}^2\right)
	F_1\left(\frac{3}{4};\frac{1}{2},\frac{1}{2};\frac{7}{4};-\frac{r^4 \beta ^2}{q_{E}^2},-\frac{r^4 \beta ^2}{q_{M}^2}\right)\right)}{21 q_{E} q_{M}} 
. \label{MM}
\end{eqnarray}

Finally, by replacing the geometric mass with Eq. \eqref{MM} in
$T = \frac{1}{4 \pi }(\frac{d \psi (r)}{dr})|_{r=r_{+}}$, we find the
temperature as

\begin{eqnarray}
T & = &  \frac{k r_{+}^2 - r_{+}^4 (\Lambda-2 \beta ^2)- 2
	\left( \beta ^2 r_{+}^4 +  q_{E}^2 \right) \sqrt{\frac{q_{M}^2+\beta ^2 r_{+}^4}{q_{E}^2+\beta ^2 r_{+}^4}}}{4
	\pi  r_{+}^3}. \label{temp}
\end{eqnarray}

The positivity of the temperature is one of the physical restrictions on
our solutions. To meet this end, we investigate the limiting behaviors
of the temperature as

\begin{equation}
\lim_{r_{+}\longrightarrow 0 }T=-\frac{	q_{M}q_{E}}{2 \pi r_{+}^3}+\frac{ k}{4 \pi r_{+}} +O(r_{+}),
\end{equation}
\begin{equation}
\lim_{r_{+}\longrightarrow \infty }T=-\frac{ \Lambda r_{+}}{ 4 \pi}+\frac{ k}{4 \pi r_{+}}+O(\frac{1}{r_{+}^2}),
\end{equation}
where the ${r_{+} } \to 0$ is related to high energy limit and ${r_{+} } \to \infty $ investigates the asymptotic behavior of the temperature. For $AdS$ black holes: I) At least one root exists for temperature. II) For small (large) black holes, temperature is always
negative (positive) and it is governed by matter (gravitational) field
contributions. Medium black holes's behaviors are determined by
topological factor.

For $dS$ black holes: a) Temperature could be completely negative (for
hyperbolic horizon and $\Lambda \leq 2 \beta ^{2}$), with one root, or
with two roots (for spherical black holes). b) Small and large black
holes have negative temperature, while the medium black hole's
temperature is determined by horizon of the black holes.

Next thermodynamical quantity of interest is the total mass of black
holes. To calculate this property, one can use Arnowitt-Deser-Misner
approach \cite{ADM} which yields

\begin{eqnarray}
M & = &  \frac{m}{8 \pi}. 
\end{eqnarray}

Using obtained geometrical mass \eqref{MM}, we find the mass as

\begin{eqnarray}
M & = & \frac{k r_{+}}{8 \pi }+\frac{q_{E}^2 \sqrt{\frac{q_{M}^2+\beta ^2
			r_{+}^4}{q_{E}^2+\beta ^2 r_{+}^4}}}{4 \pi  r_{+}}-\frac{(\Lambda-2 \beta ^2)  r_{+}^3}{24 \pi } -\frac{\beta ^4 r_{+}^7 F_1\left(\frac{7}{4};\frac{1}{2},\frac{1}{2};\frac{11}{4};-\frac{r_{+}^4 \beta
		^2}{q_{E}^2},-\frac{r_{+}^4 \beta ^2}{q_{M}^2}\right)}{7 \pi  q_{E} q_{M}}+                                    \notag
\\[0.1cm] 
&  &
\frac{\beta ^2 r_{+}^3 \left(3q_{E} q_{M} \sqrt{\frac{q_{M}^2+\beta
			^2 r_{+}^4}{q_{E}^2+\beta ^2 r_{+}^4}}-2 \left(q_{E}^2+q_{M}^2\right)
	F_1\left(\frac{3}{4};\frac{1}{2},\frac{1}{2};\frac{7}{4};-\frac{r_{+}^4 \beta
		^2}{q_{E}^2},-\frac{r_{+}^4 \beta ^2}{q_{M}^2}\right)\right)}{12 \pi  q_{E}q_{M}}. \label{mass}
\end{eqnarray}

In order to calculate the total electric and magnetic charges of black
holes, one can use Gauss law \cite{Hajkhalili}. For the total
electric charge, this law is given by
  
\begin{eqnarray}
Q_{E}=\frac{1}{4 \pi} \int_{r \rightarrow \infty}\sqrt{-g}F_{tr}d^{2}x,
\end{eqnarray}
which gives the total electric charge per unit volume as  

\begin{eqnarray}
Q_{E}=\frac{q_{E}}{4 \pi}. \label{QE}
\end{eqnarray}

The same method could be applied to find total magnetic charge per unit
volume as

\begin{eqnarray}
Q_{M}=\frac{q_{M}}{4 \pi}. \label{QM}
\end{eqnarray}

In order to calculate the electric and magnetic potentials, one can use
the free energy approach \cite{Dutta,Hajkhalili}. The free energy
is given by

\begin{equation}
W=\frac{I_{on \text{ } shell}}{\zeta},
\end{equation}
where $I_{on \text{ } shell}$ is on shell action and $\zeta $ is the
inverse of temperature. It is a matter of calculation to find electric
and magnetic potentials as \cite{Dutta,Hajkhalili}
 
\begin{eqnarray}
U_{E}=-\frac{dW}{dQ_{E}}=-\frac{\Omega}{21
	q_{E}^2 q_{M} r_{+}}, \label{UE}
\end{eqnarray}

\begin{eqnarray}
U_{M}=\frac{dW}{dQ_{M}}=-\frac{\Omega}{21
	q_{M}^2 q_{E} r_{+}}. \label{UM}
\end{eqnarray}
where 

\begin{eqnarray}
	\Omega & = & 7 \beta ^2 r_{+}^4 \left(2 q_{E}^2+q_{M}^2\right)
	F_1\left(\frac{3}{4};\frac{1}{2},\frac{1}{2};\frac{7}{4};-\frac{r_{+}^4 \beta
		^2}{q_{E}^2},-\frac{r_{+}^4 \beta ^2}{q_{M}^2}\right)+9 \beta ^4 r_{+}^8
	F_1\left(\frac{7}{4};\frac{1}{2},\frac{1}{2};\frac{11}{4};-\frac{r_{+}^4 \beta
		^2}{q_{E}^2},-\frac{r_{+}^4 \beta ^2}{q_{M}^2}\right)-    \notag
	\\ 
	& & 
	21 q_{E} q_{M}
	\sqrt{\left(q_{E}^2+\beta ^2 r_{+}^4\right) \left(q_{M}^2+\beta ^2 r_{+}^4\right)}.
\end{eqnarray}

Our final investigation in this section is checking the validation of
the first law of black hole thermodynamics. In our black holes as
thermodynamical systems, mass plays the role of internal energy, and
entropy, total electric and magnetic charges are extensive quantities,
whereas temperature, total electric and magnetic potentials are
intensive quantities. Therefore, the first law of black holes
thermodynamics is given by
 
\begin{equation}
dM=TdS+U_{E}dQ_{E}+U_{M}dQ_{M},  \label{first law}
\end{equation}
where 

\begin{equation}
T=\left( \frac{\partial M}{\partial S}\right) _{Q_{E},Q_{M}}\ \ \ \ \ \ \ \&\ \ \ \ \
\ \ \ U_{E}=\left( \frac{\partial M}{\partial Q_{E}}\right) _{S,Q_{M}}\ \ \ \ \ \ \ \&\ \ \ \ \
\ \ \ U_{M}=\left( \frac{\partial M}{\partial Q_{M}}\right) _{S,Q_{E}}.  \label{TU}
\end{equation}

It is a matter of calculation to show that obtained temperature \eqref{temp}, electric \eqref{UE} and magnetic \eqref{UM} potentials coincide with calculated ones using the first law of black holes
thermodynamics. This confirms the validation of first law of black hole thermodynamics.

\subsection{Thermal stability} 

In the next step, we study the contributions of nonlinear
electromagnetic field on thermal stability of the solutions. The aim is
to investigate thermodynamical phases available for the solutions, their
stability/instability and type of transition between them. To do so, we
use heat capacity obtained as

\begin{eqnarray}
C= T\frac{(\frac{dS}{dr_{+}})_{q_{M},q_{E}}}{(\frac{dT}{dr_{+}}%
	)_{q_{M},q_{E}}} & = &  \frac{k r_{+}^2 - r_{+}^4 (\Lambda-2 \beta ^2)- 2
	\left( \beta ^2 r_{+}^4 +  q_{E}^2 \right) \sqrt{\frac{q_{M}^2+\beta ^2 r_{+}^4}{q_{E}^2+\beta ^2 r_{+}^4}}}{
	2r_{+}^2 ( 2 \beta ^2 - \Lambda)- 2k + \frac{4 \left(\beta ^2 r_{+}^2 \left(q_{E}^2+q_{M}^2\right) + 3 q_{E}^2 q_{M}^2 - \beta ^4
		r_{+}^8\right)}{r_{+}^2 \left(q_{M}^2+\beta ^2 r_{+}^4\right)}\sqrt{\frac{q_{M}^2+\beta ^2 r_{+}^4}{q_{E}^2+\beta ^2 r_{+}^2}}}. \label{heat}
\end{eqnarray}

The negativity/positivity of heat capacity indicates black holes having
unstable/stable thermal phase. Usually, different phases are separated
by transition points which could be roots and/or divergencies of the
heat capacity. In the case of divergency, the transition between phases
is thermodynamical phase transition.

\begin{figure*}[!htbp]
	\centering
	\subfloat[$q_{M}=0.1$, $\beta=0.1$ and $\Lambda=-1$]{\includegraphics[width=0.25\textwidth]{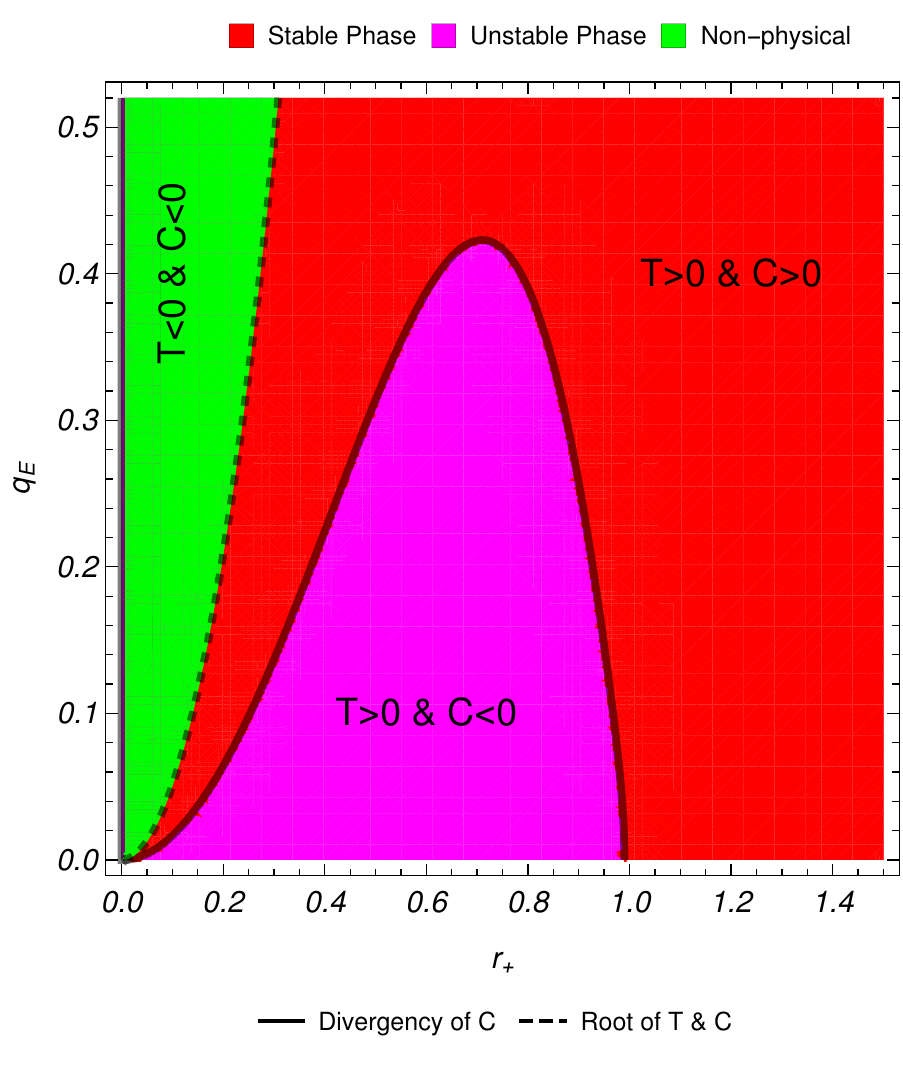} \label{HqE}}
	\subfloat[$q_{E}=0.1$, $\beta=0.1$ and $\Lambda=-1$]{\includegraphics[width=0.25\textwidth]{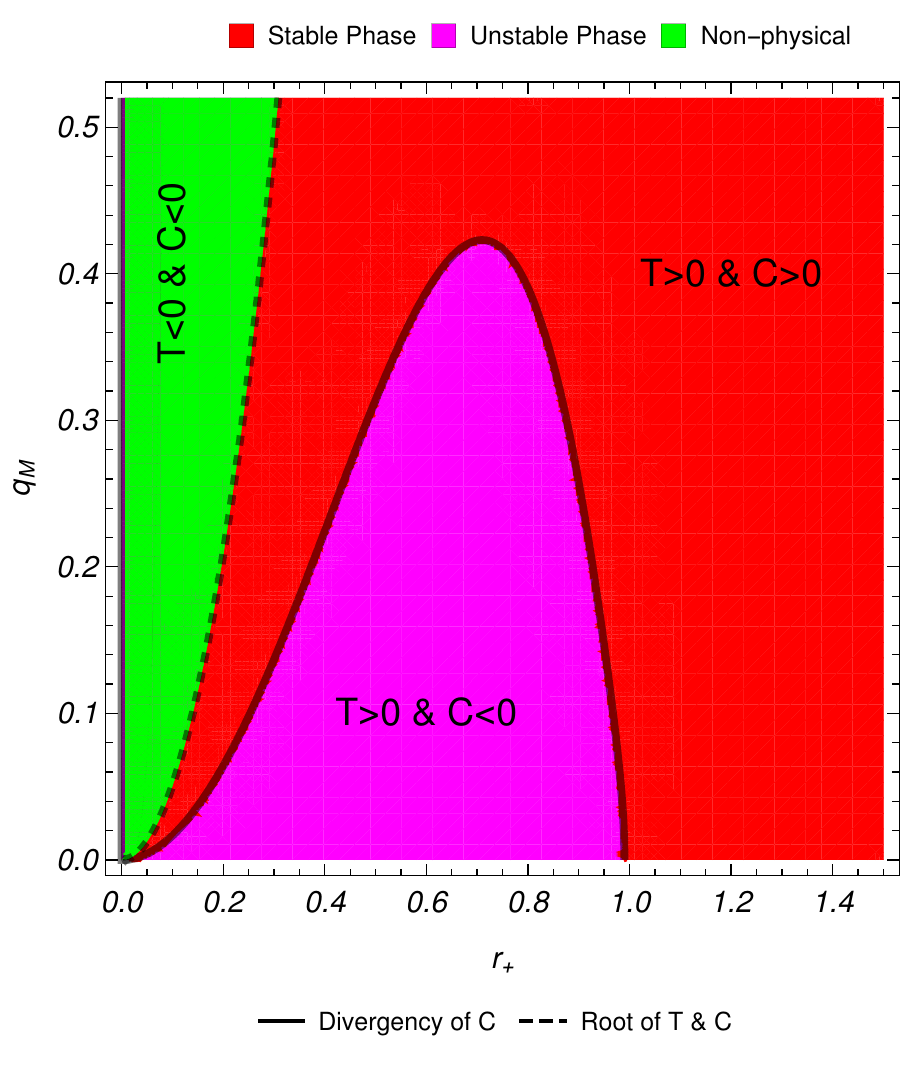} \label{HqM}}
	\subfloat[$q_{E}=0.1$, $q_{M}=0.3$ and $\Lambda=-1$]{\includegraphics[width=0.25\textwidth]{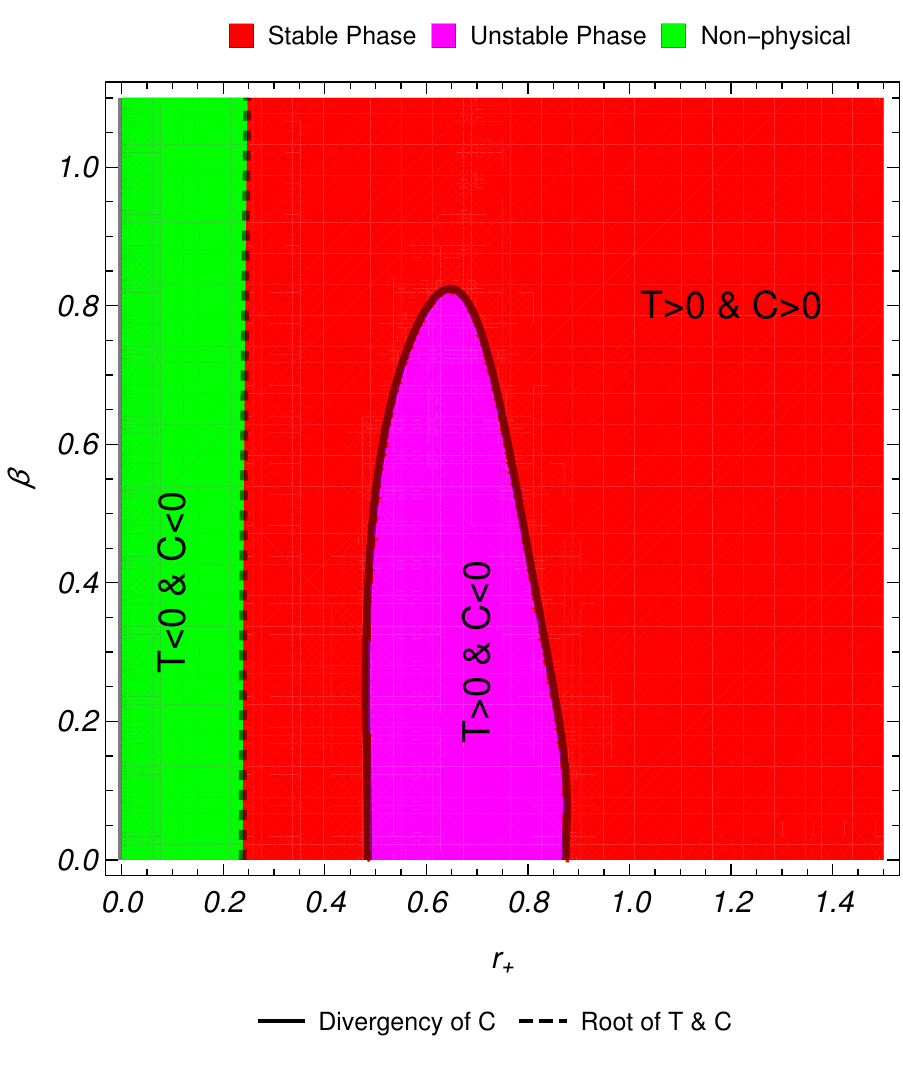} \label{Hb}}
	\subfloat[$q_{E}=0.1$, $q_{M}=0.3$ and $\beta=0.1$.]{\includegraphics[width=0.25\textwidth]{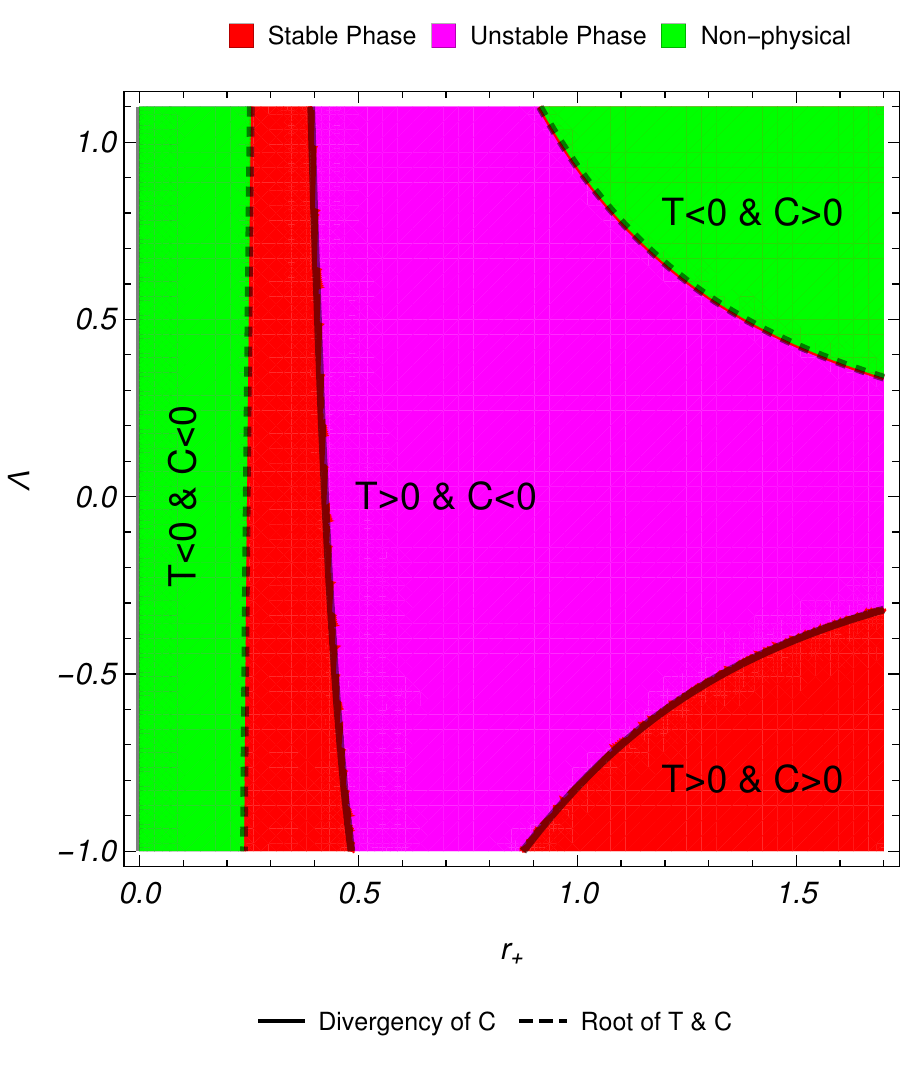} \label{HL}}
	\caption{Heat capacity versus horizon radius for $k=1$.} \label{Fig1}
\end{figure*}

The obtained heat capacity is more complex to calculate its roots and
divergencies analytically. Therefore, we use numerical method to
investigate the effects of different parameters on heat capacity in Fig.~\ref{Fig1}. In order to have a more precise picture, we have
distinguished regions with negative temperature as ``non-physical'' in
diagrams. 

Evidently, for highly charged electric (Fig.  \ref{HqE}) and magnetic (Fig. \ref{HqM}) fields and also large nonlinearity parameter (Fig. \ref{Hb}),
there are two phases available for these black holes: small black holes
which are non-physical and large thermally stable ones. These two phases
are separated by the root of temperature/heat capacity. The existence
of critical behavior is limited by an upper bound coming from
nonlinearity parameter, electric and magnetic charges. There are
specific critical values for these parameters where black holes would
have three phases: small non-physical black holes, medium and large
stable ones. The medium and large phases are separated by divergence
point which is known as critical point. Therefore, here we have a
thermal phase transition between two stable phases. For lightly
electrically and magnetically charged cases and small nonlinearity
parameter, there will be four phases: very small non-physical black
holes, small stable ones, medium unstable ones and large stable ones.
The transition between three phases of stable and unstable takes place
through divergencies, hence thermal phase transitions. These effective
behaviors are observed for $AdS$ black holes. For $dS$
case (Fig. \ref{HL}), there could be four phases available for black holes as
well. But here, very small and large black holes are non-physical,
medium black holes are unstable and only small black holes are stable.
The transition between these two physical phases takes place through a
thermal phase transition.

In conclusion, we see that if our black hole solutions move toward
situation with small electric and/or magnetic charge, it would
critically become active and several phases with specific thermal phase
transitions would be observed for them. In context of nonlinearity
parameter, small values of it indicates high strength for nonlinear
nature of the electromagnetic field. Therefore, we see for high degrees
of nonlinearity, system would acquire critical behavior. In contrast,
if electromagnetic field behaves like Reissner-Nordstr\"{o}m black holes
with dyonic charge, the critical points would vanish and thermally
stable black holes would be resulted.

\subsection{Extended phase space thermodynamics} 

Next, we consider the negative branch of cosmological constant to be a
thermodynamical quantity, pressure. Such a proposal could be used to
show the presence of van der Waals like behavior for black holes
\cite{Kubiznak}. The cosmological constant and pressure are related to
each other with following relation

\begin{equation}
\Lambda=-8\pi P.
\end{equation}

If we use this relation, we can recognize the obtained temperature \eqref{temp} as equation of state and find the pressure as

\begin{eqnarray}
P & = &  \frac{4 \pi  r_{+}^3 T -k r_{+}^2-2 \beta ^2 r_{+}^4 + (2 \beta ^2 r_{+}^4 + 2 q_{E}^2) \sqrt{\frac{q_{M}^2+\beta ^2 r_{+}^4}{q_{E}^2+\beta ^2 r_{+}^4}}}{8 \pi r_{+}^4}. \label{pressure}
\end{eqnarray}

To understand the pressure in more details, we find its limiting
behaviors as follows
\begin{equation}
\lim_{r_{+}\longrightarrow 0 }P=-\frac{	q_{M}q_{E}}{2 \pi r_{+}^4}-\frac{ k}{8 \pi r_{+}^2} +\frac{ T}{ 2 r_{+}}+\frac{	(q_{E}-q_{M})^2 \beta^2}{8 \pi q_{M}q_{E}}+O(r_{+}^3),
\end{equation}
\begin{equation}
\lim_{r_{+}\longrightarrow \infty }P=\frac{ T}{ 2 r_{+}}-\frac{ k}{4 \pi r_{+}^2}+\frac{	q_{E}^2+q_{M}^2 \beta^2}{8 \pi r_{+}^4}-\frac{	(q_{E}-q_{M})^2 }{32 \pi \beta^2 r_{+}^8}+O(\frac{1}{r_{+}^9}).
\end{equation}

Evidently, for large and small black holes, the pressure would be
positive. In contrast, for medium black holes, the pressure could vanish
and/or be negative. According to classical thermodynamics, pressure
should be positive valued and a decreasing function of the volume. The
region where pressure is an increasing function of volume is not
accessible for the thermodynamical system and thermal phase transition
takes place over it.

The volume could be obtained by extended version of first law black
holes thermodynamics given by 

\begin{equation}
dM=TdS+VdP+U_{E}dQ_{E}+U_{M}dQ_{M},  \label{first law extended}
\end{equation}
which gives us the volume as
\begin{equation}
V=(\frac{dM}{dP})_{q_{M},q_{E}}=\frac{r_{+}^3}{3}.  \label{volume}
\end{equation} 

\begin{figure*}[!htbp]
	\centering
	\subfloat[$q_{M}=0.1$, $\beta=0.1$ and $T=0.1$]{\includegraphics[width=0.25\textwidth]{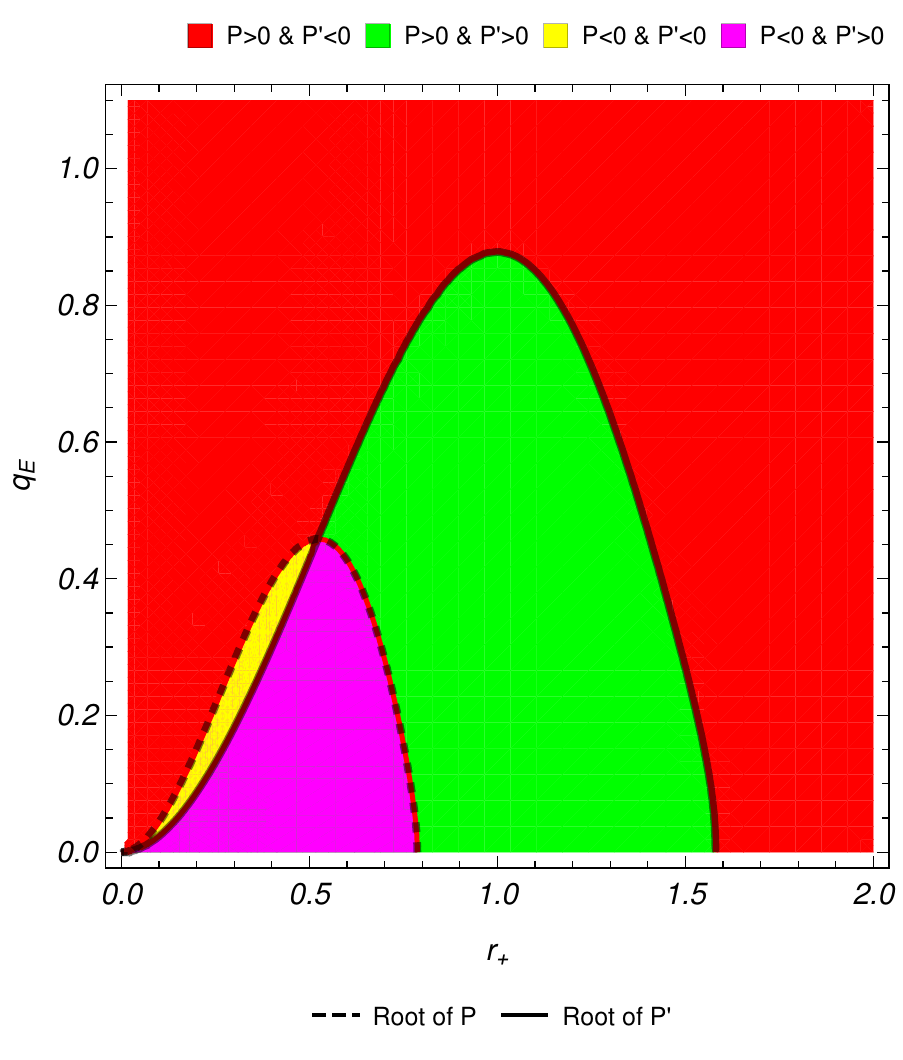} \label{PqE}}
	\subfloat[$q_{E}=0.1$, $\beta=0.1$ and $T=0.1$]{\includegraphics[width=0.25\textwidth]{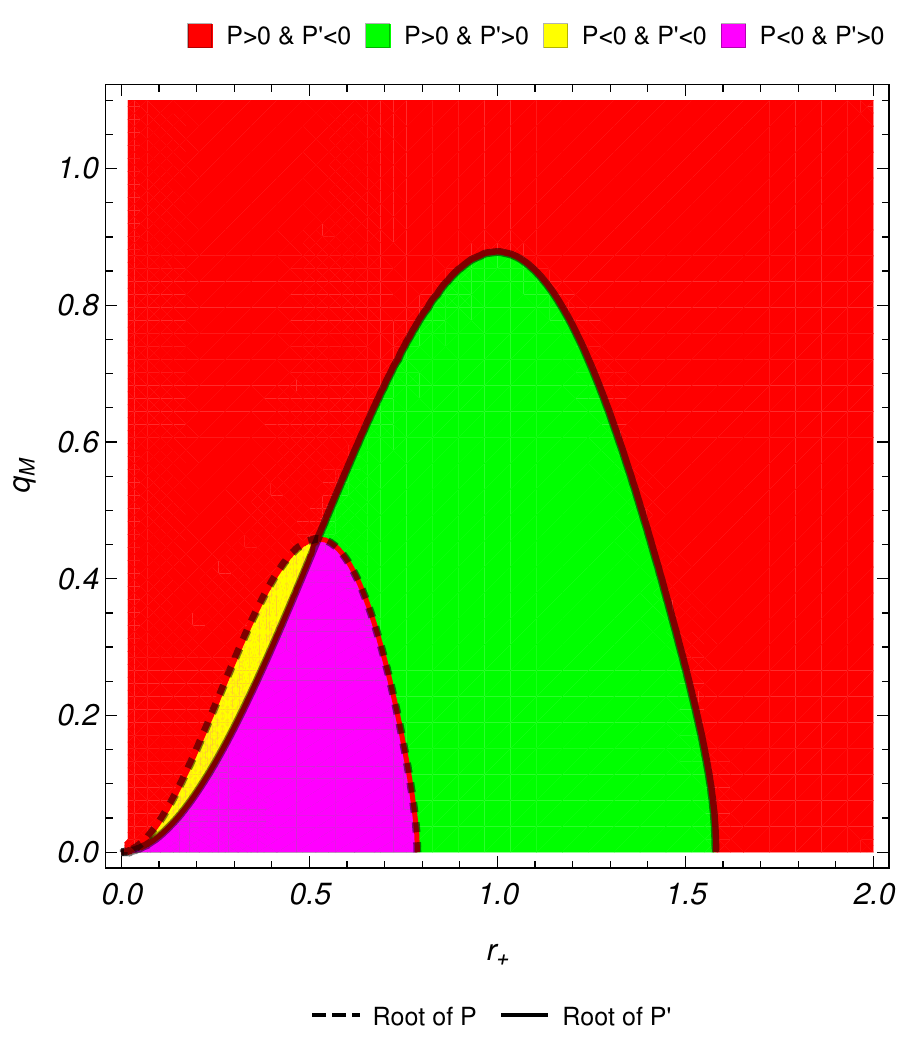} \label{PqM}}
	\subfloat[$q_{E}=0.1$, $q_{M}=0.3$ and $T=0.1$]{\includegraphics[width=0.25\textwidth]{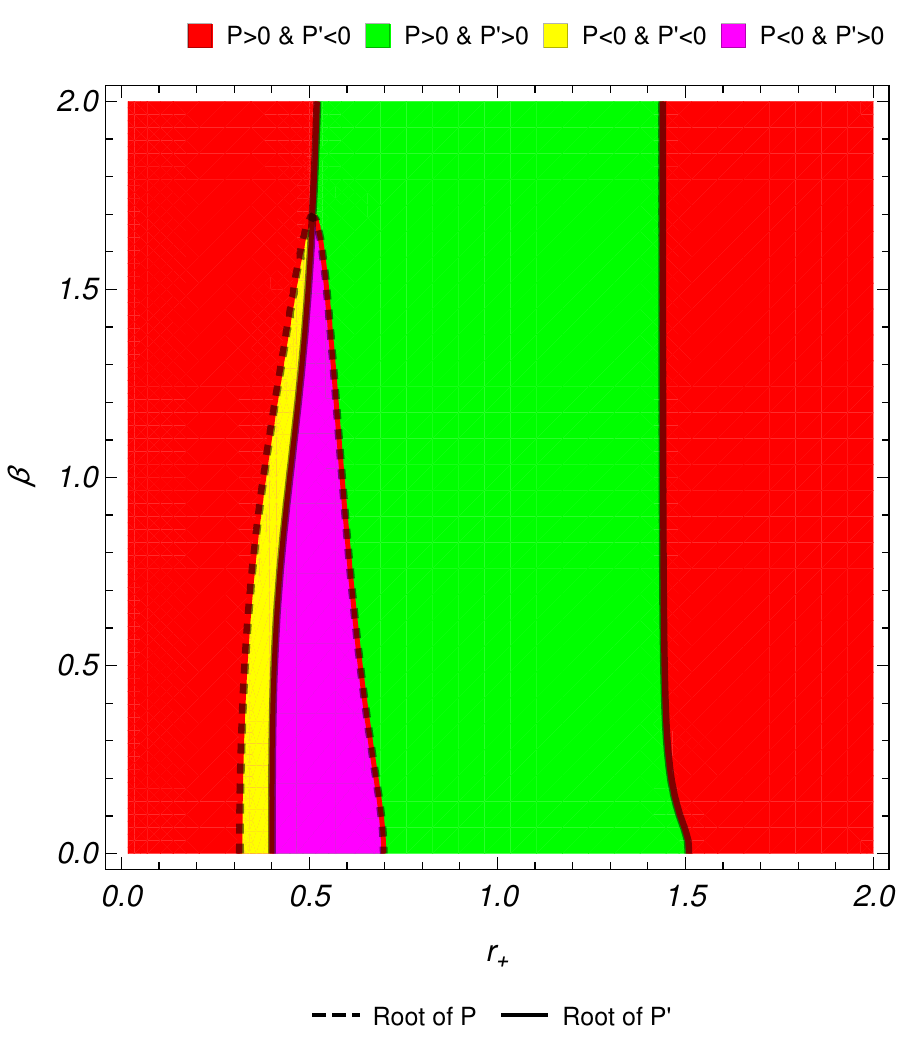} \label{Pb}}
	\subfloat[$q_{E}=0.1$, $q_{M}=0.3$ and $\beta=0.1$]{\includegraphics[width=0.25\textwidth]{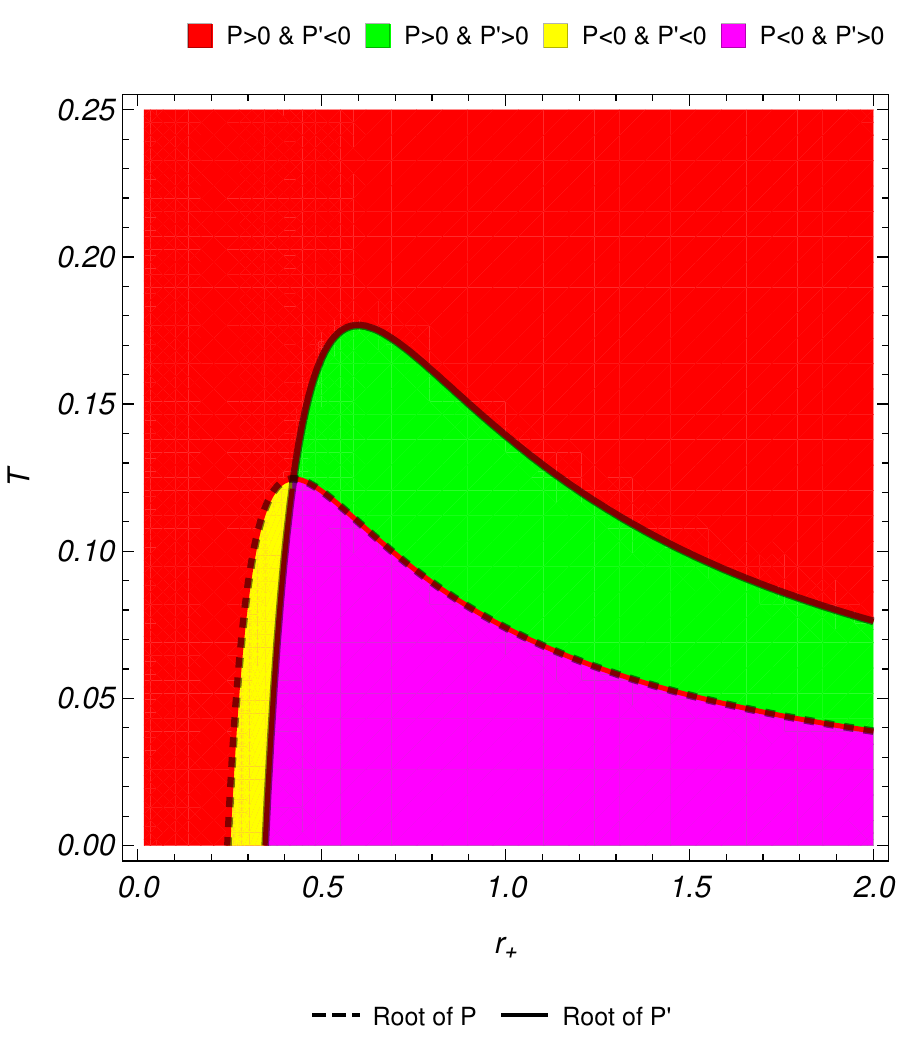} \label{PL}}
	\caption{Pressure and $P'$  versus horizon radius for $k=1$.} \label{Fig2}
\end{figure*}

Since the volume and horizon radius are cubically related to each other, one can
use the horizon radius instead of volume for investigating
thermodynamical properties of black holes. If one wants pressure to be
a decreasing function of the volume (horizon radius), the derivation of
pressure with respect to horizon radius should be negative ($P'\equiv
\frac{dP}{dr_{+}}<0$). To have a full picture regarding the effects of
different parameters on behavior of the pressure and distinguish
physical regions, we have plotted the following diagrams in Fig.~\ref{Fig2}.

%
%
%
%

Thermodynamically acceptable behaviors are where $P>0$ and $P'<0$.
Accordingly, the roots of pressure and $P'$ could be used as starting
points to describe (non)physical regions and the effects of different
parameters on them.

Evidently, for high temperature, nonlinearity parameter, electric and
magnetic charged, black holes's pressure is positive while $P'$ is
negative. Therefore, for any volume (horizon radius), black hole's
pressure has a sound behavior, hence solutions are physical. In
contrast, for small values of these parameters, pressure and $P'$ could
acquire first one and then two roots. This indicates that for small
temperature, nonlinearity parameter, electric and magnetic charged,
black holes develop regions in which pressure could be negative and/or
$P'$ is positive. Since these are physically forbidden, one can draw the
conclusion that these regions do not admit black hole solutions. In
case where both pressure and $P'$ have two roots, the nonphysical region
is determined by smaller root of the pressure and larger root of $P'$.
Interestingly, when pressure has one root and $P'$ have two roots, the
root of the pressure coincide with smaller root of $P'$.

The observed behaviors confirm that: For Maxwell like electromagnetic
field (large nonlinearity parameter), there is no bound on the pressure
as a function of the volume. In contrast, for highly nonlinear regime,
pressures behavior would be bounded by conditions coming from its sign
and being a decreasing function of volume. From electric and magnetic
charge perspectives, restrictive behavior for pressure is observed for
lightly charged cases. The super magnetic and electric charged black
holes would have no restriction on their pressures. Surprisingly, the
bounded behavior for pressure could be seen for cold black holes while
the hot ones prove to have no limitation on their pressure.

It should be noted, one can write the equation of state \eqref{pressure} in form of thermodynamical quantities as 

\begin{eqnarray}
P & = & \frac{32 \pi  S^{3/2} T-4 k S-32 \beta ^2 S^2}{128 \pi  S^2}+\frac{\left(32 \pi ^2 Q_{E}^2+32 \beta ^2 S^2\right) \sqrt{\frac{16 \pi ^2 Q_{M}^2+16 \beta ^2 S^2}{16 \pi ^2 Q_{E}^2+16 \beta ^2 S^2}}}{128
	\pi  S^2}, \label{PP1}
\end{eqnarray}
or replace the entropy with volume and find
\begin{eqnarray}
P & = & \frac{12 \pi  T V-3^{2/3} k V^{2/3}-6 \sqrt[3]{3} \beta ^2 V^{4/3}}{24 \sqrt[3]{3} \pi  V^{4/3}}+\frac{\left(32 \pi ^2 Q_{E}^2+6 \sqrt[3]{3} \beta ^2 V^{4/3}\right) \sqrt{\frac{16 \pi ^2 Q_{M}^2+3
			\sqrt[3]{3} \beta ^2 V^{4/3}}{16 \pi ^2 Q_{E}^2+3 \sqrt[3]{3} \beta ^2 V^{4/3}}}}{24 \sqrt[3]{3} \pi  V^{4/3}}, \label{PP2}
\end{eqnarray}
in any case, the discussion regarding the effects of different
parameters on thermodynamical behavior of the system would yield results
whether one uses Eq. \eqref{pressure} or Eqs. \eqref{PP1} and/or \eqref{PP2}.

In next step, we investigate the type of thermal phase transition that
these black holes could have. To do so, we use the equation of state \eqref{pressure} and the concept of the inflection point given by 

\begin{equation}
\left( \frac{\partial P}{\partial r_{+}}\right)_{T,q_{M},q_{E}} =\left( 
\frac{\partial ^{2}P}{\partial r_{+}^{2}}\right)_{T,q_{M},q_{E}} =0,
\label{infel}
\end{equation}
where it would yield the following equation for calculating critical
horizon radius (volume)

\begin{equation}
\frac{k r_{+}^2-\frac{2 \left(\frac{q_{M}^2+\beta ^2 r_{+}^4}{q_{E}^2+\beta ^2 r_{+}^4}\right)^{3/2}
		\left(6 q_{E}^4 q_{M}^4+3 \beta ^6 r_{+}^{12} \left(q_{E}^2+q_{M}^2\right)+9 \beta ^2
		q_{E}^2 q_{M}^2 r_{+}^4 \left(q_{E}^2+q_{M}^2\right)+\beta ^4 r_{+}^8 \left(q_{E}^4+16
		q_{E}^2 q_{M}^2+q_{M}^4\right)\right)}{\left(q_{M}^2+\beta ^2 r_{+}^4\right)^3}}{4 \pi
	r_{+}^3}=0.
\label{critical expression}
\end{equation}

\begin{figure*}[!htbp]
	\centering
	\subfloat[Pressure versus horizon radius; Bold line: $T=0.176$; Dashed line: The area under it, is forbidden.]{\includegraphics[width=0.3\textwidth]{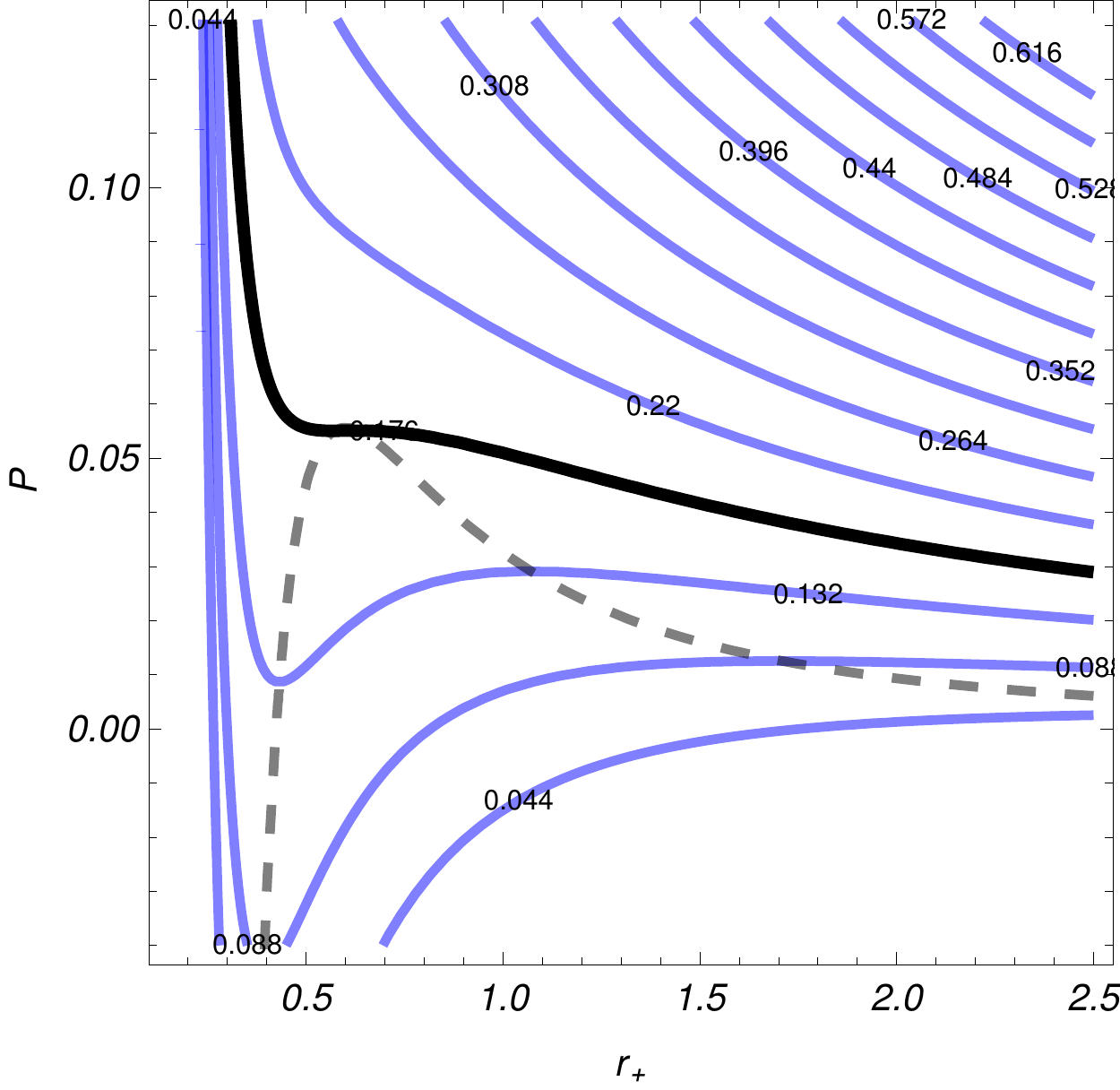} \label{Pr}}
	\subfloat[Temperature versus horizon radius; Bold line: $P=0.0558$; Dashed line: The area under it, is forbidden.]{\includegraphics[width=0.3\textwidth]{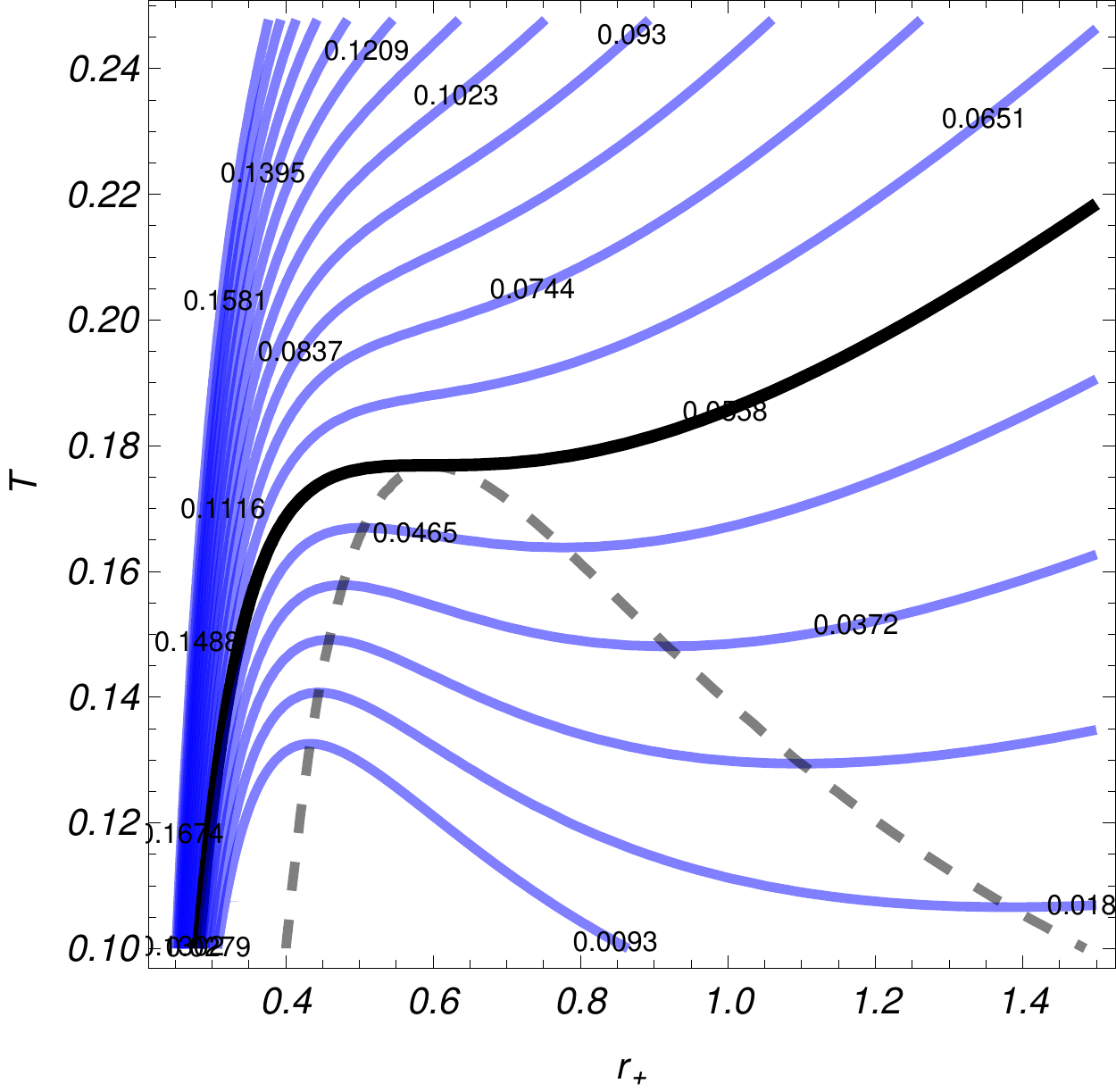} \label{Tr}}
	\caption{Critical behavior for $q_{M}=3$, $q_{E}=0.1$, $k=1$ and $\beta=0.1$.} \label{Fig3}
\end{figure*}

Due to complexity of this relation, it is not possible to obtain
critical horizon radius analytically, therefore, we use numerical
method. As an example, we have plotted diagrams in Fig.~\ref{Fig3}. The
presences of subcritical isothermal bars and isobars show that the type
of phase transition is van der Waals like. For pressures larger than
critical pressure, black holes have uniform phases without any phase
transitions. In contrast, when the pressure of the black holes becomes
smaller than critical pressure, black holes develop a region where
physical behavior is not observed. Therefore, a phase transition would
take place between two different volumes (horizon radius) with same
pressure (see Fig.~\ref{Pr}). The same could be stated for
temperature as well (see Fig.~\ref{Tr}). These are the
characteristics of van der Waals like phase transition of liquid-gas.

\begin{figure*}[!htbp]
	\centering
	\subfloat[$q_{E}=0.1$.]{\includegraphics[width=0.25\textwidth]{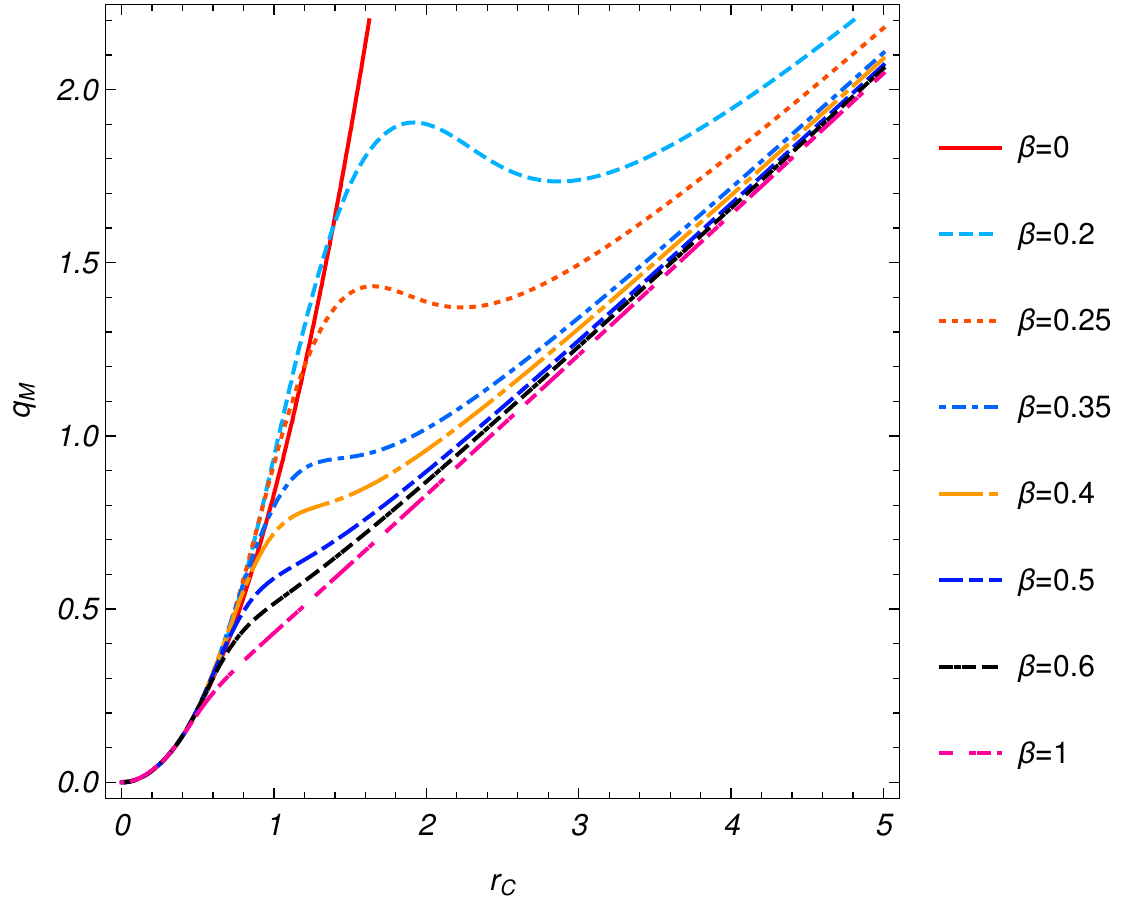} \label{rc2}}
	\subfloat[$\beta=0.1$.]{\includegraphics[width=0.25\textwidth]{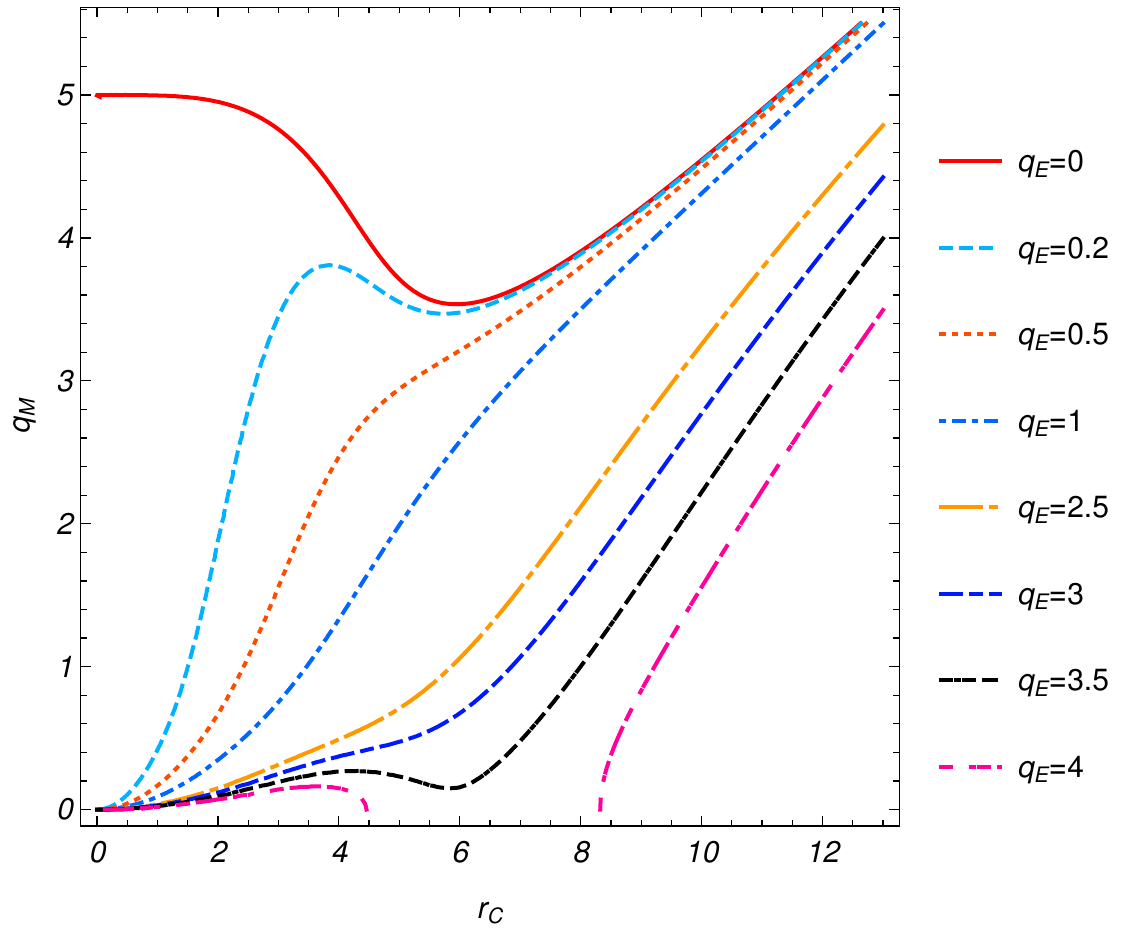} \label{rc1}}
	\subfloat[$q_{E}=0.1$.]{\includegraphics[width=0.25\textwidth]{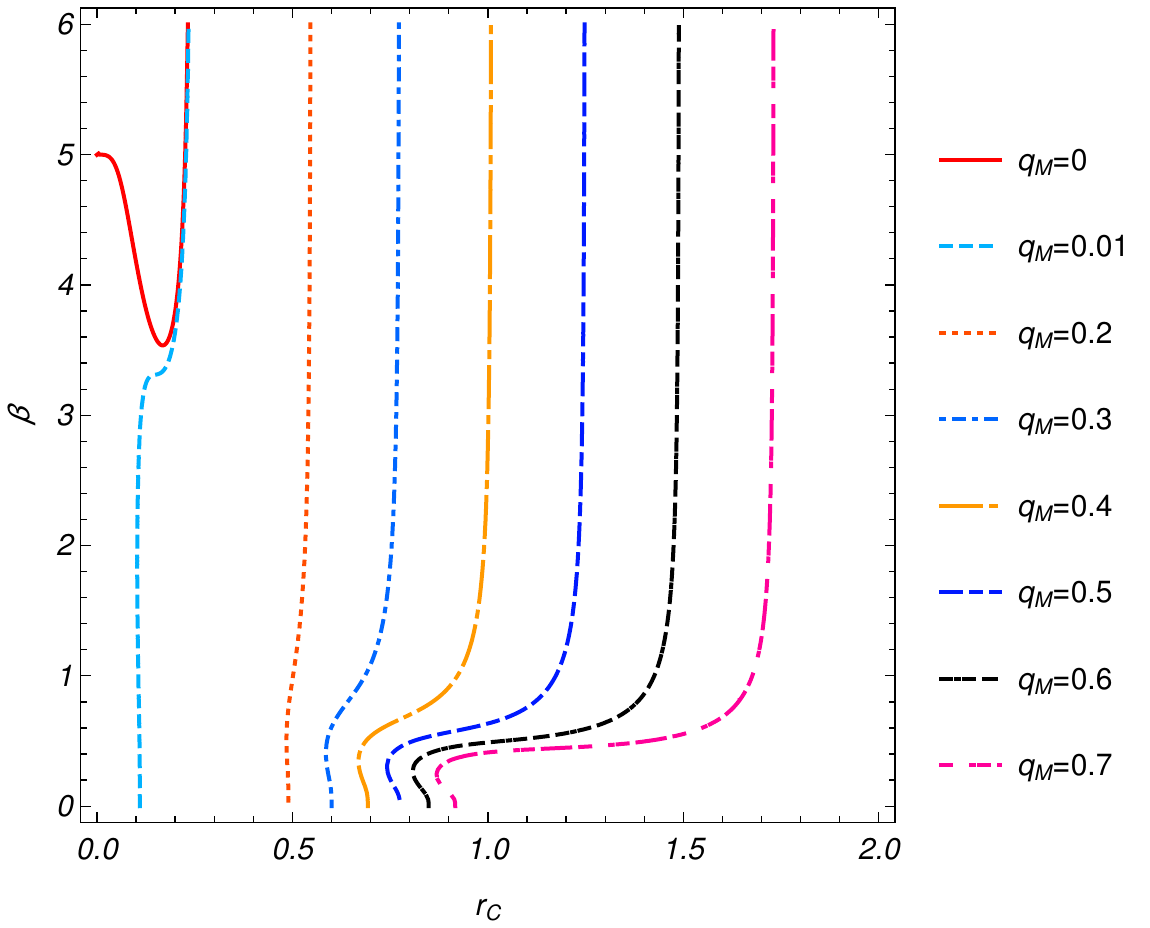} \label{rc3}}
	\subfloat[$q_{E}=0.1$.]{\includegraphics[width=0.25\textwidth]{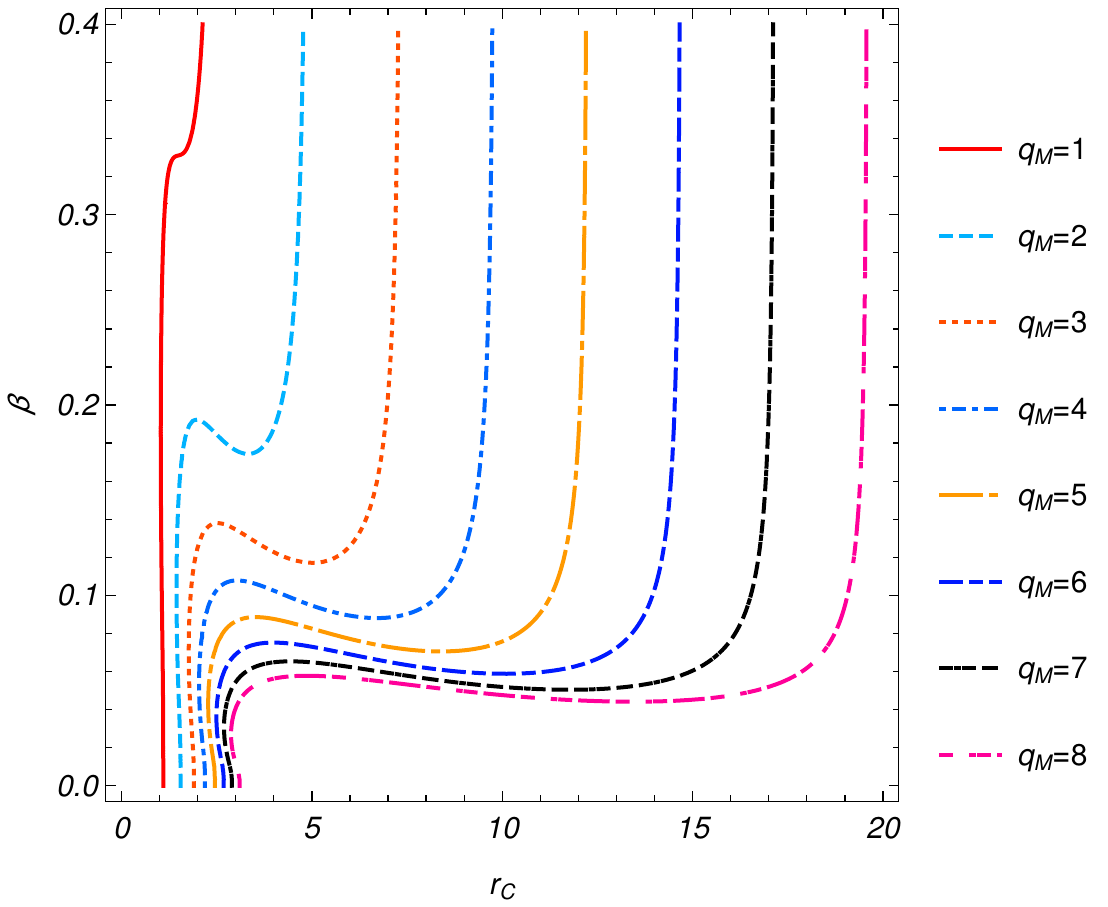} \label{rc4}}
	\caption{Variation of critical horizon radius as a function of different parameters with $k=1$. In panels \ref{rc2}, \ref{rc3} and \ref{rc4}, the electric charge is fixed and the effects of variations of nonlinearity parameter and magnetic charge on critical horizon are investigated.} \label{Fig4}
\end{figure*}

To understand the critical behavior of the system in more details, we
use numerical method to obtain critical horizon radius as a function of
different parameters. The results are given in Fig.~\ref{Fig4}.

The following points could be understood from obtained diagrams:

I) Irrespective of choices for different parameters, for large magnetic
charge and nonlinearity parameter, black holes have only one critical
horizon radius. This shows that for highly magnetized solutions and
Maxwell-like case, there is only one critical point where the usual van
der Waals like phase transition would take place.

II) In the absence of nonlinearity parameter (Fig. \ref{rc2}), the critical
horizon radius has an exponential growth as a function of magnetic
charge. In contrast, for large nonlinearity parameter, it would be a
linear growth. For the medium nonlinearity parameter, there are two
specific magnetic charges ($q_{M_{1}}$ and $q_{M_{2}}$) where for
$q_{M_{1}}\leq q_{M} \leq q_{M_{2}}$, there would be two or three
critical horizon radii for the same magnetic charge. This indicates
that for specific values of different parameters, our critical equations
yield two or three distinguishable critical horizon radii, temperatures
and pressures. This is a phenomena in thermally critical systems known
as the reentrant of phase transition.

III) In the absence of electric charge (Fig. \ref{rc1}), there is a minimum
for magnetic charge ($q_{M_{min}}$) where for $q_{M} < q_{M_{min}}$,
there is no critical horizon radius, hence no critical behavior. In
contrast, for $q_{M_{min}}\leq q_{M} \leq q_{M_{f}}$, there would be two
critical horizon radii for each magnetic charge and finally in case of
$ q_{M_{f}} < q_{M}$, the critical horizon radius has a linear growth
as a function of magnetic charge. For medium electric charge, the
critical horizon radius has a cubic growth as a function of magnetic
charge. Interestingly, for small and large electric charges, similar to
the medium nonlinearity parameter case, there are two specific magnetic
charges ($q_{M_{3}}$ and $q_{M_{4}}$) where for $q_{M_{3}}\leq q_{M}
\leq q_{M_{4}}$, there would be two or three critical horizon radii for
the same magnetic charge.

IV) In the absence of magnetic charge (Fig. \ref{rc3}), there is a minimum
for nonlinearity parameter ($\beta _{min}$) where for $\beta < \beta
_{min}$, there is no critical horizon radius, hence no critical
behavior. In contrast, for $\beta _{min}\leq \beta \leq \beta _{f}$, there
would be two critical horizon radii for each nonlinearity parameter and
finally in case of $ \beta _{f} < \beta $, the critical horizon radius
has a linear growth as a function of nonlinearity parameter. For small
magnetic charge, for each nonlinearity parameter, there is one critical
horizon radius. In case of medium magnetic charge, there are two
critical horizon radii, ($r_{C_{1}}$ and $r_{C_{2}}$) where for
$r_{C_{1}}\leq r_{C} \leq r_{C_{2}}$, there are two values of
nonlinearity parameter which yield same critical horizon radius. This
is the reminiscence of phenomena known as triple point. It should be
noted that such values of nonlinear parameter are small, hence they are
in high nonlinearity regime. Finally, for large values of the magnetic
charge (Fig. \ref{rc4}), there are two specific nonlinear parameters ($\beta
_{1}$ and $\beta _{2}$) where for $\beta _{1}\leq \beta \leq \beta _{2}$,
there would be two or three critical horizon radii for nonlinear
parameter. Here as well, $\beta _{1}$ and $\beta _{2}$ are small,
therefore, such case takes place in high nonlinearity regime.

In conclusion, we observe that the black holes under consideration here
could have a thermodynamical phenomena known as the reentrant of phase
transition. As we noticed, such phenomena is taking place in
\textit{a)} high nonlinearity regime, \textit{b)} large
magnetic/electric charge, \textit{c)} absence of magnetic/electric
charge and \textit{d)} small electric charge. In addition, we observed
that black holes could have another thermodynamical phenomena known as
triple point. It should be noted that such point was observed in high
nonlinearity regime and medium magnetic charge.

\section{Reissner-Nordstr\"{o}m like behavior}

Our final discussion in this paper is reporting on an interesting phenomena
observed for these black holes. Through our investigation, we made
distinction between electric and magnetic charges, and we did not assume
that they would have identical values. If one consider electric and
magnetic charge to be identical ($q_{M}=q_{E}$), the metric function \eqref{fr} would reduce to

\begin{equation}
\psi(r) = k-\frac{m}{r}+\frac{2 q_{M}^2}{r^2}-\frac{\Lambda  r^2}{3}. \label{F1}
\end{equation}

Accordingly, if one applies the $q_{M}=q_{E}$ to temperature \eqref{temp} and mass \eqref{mass}, the following temperature and mass would be obtained, respectively

\begin{equation}
T = -\frac{-k r_{+}^2+2 q_{E}^2+\Lambda  r_{+}^4}{4 \pi  r_{+}^3}, \label{T1}
\end{equation}

\begin{equation}
M = \frac{3 k r_{+}^2+6 q_{E}^2-\Lambda  r_{+}^4}{24 \pi  r_{+}}. \label{M1}
\end{equation}

Based on the calculated temperature \eqref{T1} and entropy \eqref{entropy}, the heat capacity would be

\begin{equation}
C = \frac{-k r_{+}^4+2 q_{E}^2 r_{+}^2+\Lambda  r_{+}^6}{2 \left(k r_{+}^2-6 q_{E}^2+\Lambda  r_{+}^4\right)}, \label{C1}
\end{equation}
and pressure is given by 

\begin{equation}
P = \frac{-k r_{+}^2+2 q_{E}^2+4 \pi  r_{+}^3 T}{8 \pi  r_{+}^4}. \label{P1}
\end{equation}

The first noticeable issue is the absences of nonlinear parameter and
traces of nonlinearity behavior in metric function and thermodynamical
quantities. In fact, the obtained metric function and thermodynamical
quantities are almost identical to those calculated for
Reissner-Nordstr\"{o}m black holes with a different factor for
magnetic/electric charge. Careful examination of the factor of
magnetic/electric charge shows that we can break it down into
$q_{M}+q_{E}=2q_{E}$. Therefore, we see that this is indeed
Reissner-Nordstr\"{o}m with a magnetic charge (linearly charged dyonic
solutions). Usually, in order to remove the effects of nonlinearity, one
should consider $r \longrightarrow \infty $ or $\beta \longrightarrow
\infty $. But here, we see that due to presence of the magnetic charge
and resultant nonlinear form, we have yet another limit for obtaining
the Maxwell like behavior. The main issue is that such limit is
independent from nonlinearity parameter and $r$, and it only depends on
the amount of magnetic and electric charges. This shows that if magnetic
and electric charges become identical in these black holes, although in
essence electromagnetic field has nonlinear nature, its nonlinearity
would be totally screened by geometrical/thermodynamical
structure/properties of the black holes and they would behave like
linearly charged dyonic black holes. The effective nonlinear nature of
the electromagnetic field becomes evident only when these black holes
are charged magnetically and electrically different.

\section{Conclusion}

In this paper, we investigated nonlinearly charged dyonic black holes
in the presence of Born-Infeld electromagnetic field. The metric
function was obtained and it was shown that these black holes have an
irremovable singularity located at the origin and asymptotically
$(A)dS$ behavior. The thermodynamical quantities and behavior were
obtained and studied.

In dyonic black holes with Maxwell field, solutions has electric-magnetic
duality. Here, we showed that generalization to Born-Infeld omits such
duality. In addition, it was shown that it is possible to cancel the
effects of cosmological constant with nonlinearity parameter. In
addition, we showed that in highly electric/magnetic charged regimes and
Maxwell limits (small nonlinearity), black holes would develop only one
thermal stable phase. In contrast, for small electric/magnetic charge
and high nonlinearity regime, black holes become critically active and
several phases were observed. In studying the pressure, we showed that
in case of cold black holes, the pressure is bounded by some constraints
on its volume whereas, hot black holes' pressure has physical behavior
for any given volume.

The critical behavior of these black holes showed that generalization
to Born-Infeld would enrich phase space of these black hole and
introduce phenomena absent for Maxwell case. Among these phenomena, we
reported on possibility of existences of triple point, reentrant of
phase transition and van der Waals like behavior.

Finally, we showed that if electric and magnetic charges are identical,
the effects of the nonlinearity would be completely removed and our
solutions would behave like linearly charged dyonic black holes
(Maxwell like ones). This indicated that nonlinearity of electromagnetic
field has effective behavior which becomes evident only when these
black holes are charged magnetically and electrically different.

In the next step, one can employ the obtained results here to
investigate the AdS/CFT duality. Specially, one can focus on
diamagnetic/paramagnetic behavior on the boundary of the solutions and
modification of the Curie law due to nonlinear electromagnetic field.
In addition, it would be interesting to see whether it is possible to
see the effects of nonlinear electromagnetic field, if electric and
magnetic charges are identical, through other properties of the black
holes not investigated in this paper.

\end{document}